\documentclass[reprint,nofootinbib,amsmath,amssymb,fontenc aps]{revtex4-1}
\usepackage{graphicx,dcolumn,bm,hyperref,float}
\usepackage{anyfontsize}
\usepackage{color}

\newcommand{\be}{\begin{equation}}
\newcommand{\ee}{\end{equation}}
\newcommand{\beq}{\begin{equation}}
\newcommand{\eeq}{\end{equation}}
\newcommand{\ba}{\begin{eqnarray}}
\newcommand{\ea}{\end{eqnarray}}
\newcommand{\bef}{\begin{figure}}
\newcommand{\eef}{\end{figure}}

\newcommand{\Tr}{T_{\mbox{\tiny{reh}}}}
\newcommand{\Rbub}{R_{\text{bub}}}
\newcommand{\phiamp}{\phi_{\text{amp}}}
\newcommand{\Mpl}{M_\text{Pl}}
\newcommand{\kphys}{k_{\text{phys}}}
\newcommand{\hbub}{h_\text{bub}}
\newcommand{\hmax}{h_\text{max}}
\newcommand{\Tds}{T_\text{dS}}
\newcommand{\Hinf}{H_\text{inf}}
\newcommand{\phiinf}{\phi_\text{inf}}
\newcommand{\tf}{\text{free}}

\begin{document}

\title{Constraints on Inflaton Higgs Field Couplings}

\author{Jessie Yang$^1$}
\email{jyang58@uw.edu}
\author{Mark P.~Hertzberg$^2$}
\email{mark.hertzberg@tufts.edu}
\affiliation{$^1$Department of Physics, University of Washington, Seattle WA 98195-1560, USA
\looseness=-1}
\affiliation{$^2$Institute of Cosmology, Department of Physics and Astronomy, Tufts University, Medford, MA 02155, USA
\looseness=-1}

\begin{abstract}
According to the best-fit parameters of the Standard Model, the Higgs field's potential reaches a maximum at a field value $h \sim 10^{10-11}$\,GeV and then turns over to negative values. During reheating after inflation, resonance between the inflaton and the Higgs can cause the Higgs to fluctuate past this maximum and run down the dangerous side of the potential if these fields couple too strongly. In this paper, we place constraints on the inflaton-Higgs couplings such that the probability of the Higgs entering the unstable regime during reheating is small. To do so, the equations of motion are approximately solved semi-analytically, then solved fully numerically. Next the growth in variance is used to determine the parameter space for $\kappa$ and $\alpha$, the coupling coefficients for inflaton-Higgs cubic and quartic interactions, respectively. We find the upper bounds of $\kappa < 1.6\times 10^{-5}m_\phi\sim 2.2 \times 10^8$\,GeV and $\alpha < 10^{-8}$ to allow the Higgs to remain stable in most Hubble patches during reheating, and we also find the full two parameter joint constraints. We find a corresponding bound on the reheat temperature of $T_\text{reh} \lesssim 9.2 \times 10^9$\,GeV. Additionally, de Sitter temperature fluctuations during inflation put a lower bound on inflaton-Higgs coupling by providing an effective mass for the Higgs, pushing back its hilltop during inflation. These additional constraints provide a lower bound on $\alpha$, while $\kappa$ must also be non-zero for the inflaton to decay efficiently. 
\end{abstract}

\maketitle

\tableofcontents

\section{Introduction}

The confirmation of the Higgs particle at the LHC means that for the first time, we have a unitary theory of particle physics - the Standard Model (SM). Combined with the graviton, the theory appears to have internal consistency down to the Planck scale. However, the existence of the Higgs, the first discovered scalar in nature, opens up a new type of stability problem; as we now recap. 
The dimension 4 Higgs potential is given by
\begin{equation} \label{eq:higgspot}
V_h = -{\frac{\mu_h}{2}} H^\dagger H+{\frac{\lambda}{4}}(H^\dagger H)^2
\end{equation}
where $H$ is the Higgs doublet. This work will focus on the magnitude of $H$ which will be denoted as the Higgs field, $h$, while its angular components are reorganized into the longitudinal modes of the $W^{\pm}$ and $Z$ bosons, and will not play a direct role for us here.
As is well known, this potential has a non-zero local minimum at 246 GeV, which is classically stable and denotes our vacuum.

\begin{figure}[t]
    \centering
        \includegraphics[width=0.97\linewidth]{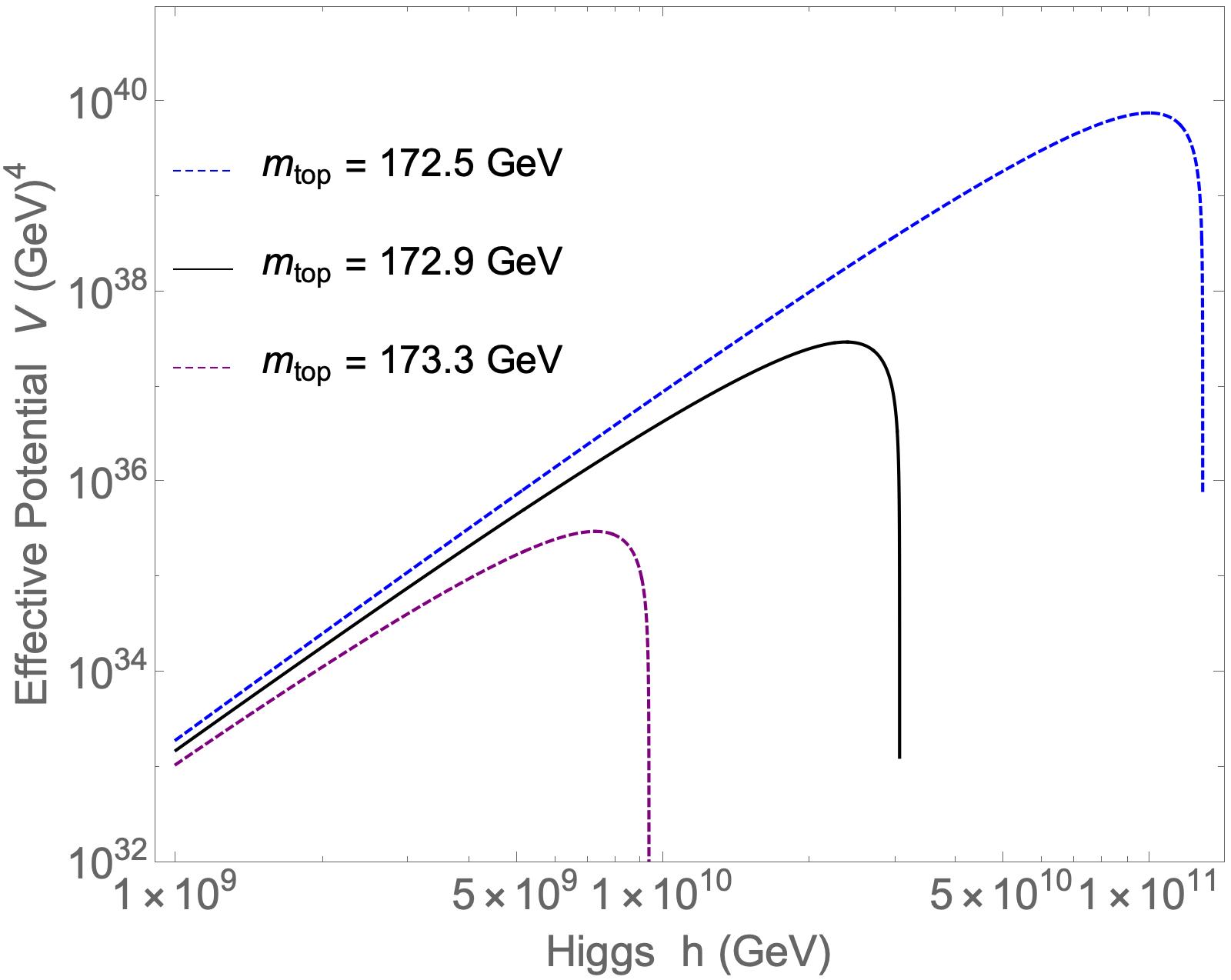}
    \caption{
   The two loop renormalized effective potential of the Higgs within the SM. The turnover at high scales around $\hmax$ ($V_h'(\hmax)=0$) is focussed on. Blue-dashed curve is $m_{top}=172.5$\,GeV, black-solid curve is $m_{top}=172.9$\,GeV, and purple-dashed curve is $m_{top}=173.3$\,GeV. After the turn-over, the potential runs negative.}
    \label{fig:higgsaccurate}
\end{figure}

At high energies, the Higgs interactions with various SM particles causes the value of the self coupling $\lambda$ to evolve under renormalization group equations. For the central values of SM parameters, this causes $\lambda$ to turn negative at large values of the Higgs field \cite{Sher:1988mj,EWvevmetastbl.2,EWvevmetastbl.3,EWvevmetastbl.1,EWvevmetastbl.4}. This means the potential reaches a maximum at some large value of the Higgs field, $\hmax$, and then quickly drop to negative values.
Figure \ref{fig:higgsaccurate} provides a plot of the effective potential from solving the two-loop renormalization group equations. This shows that the peak occurs at energies $\sim 10^{10-11}$\,GeV, depending sensitively on the top quark mass. Here we plot in black the central top mass, while the standard deviation values are given by the dashed curves \cite{PDB,Butenschoen:2016lpz}. (Note that the initial dip associated with the usual electroweak vacuum is not directly relevant as the $\mu_h h^2/2$ term in Eq.~\ref{eq:higgspot} is negligible at such high energies). The huge fluctuation required for the Higgs field to go over, or tunnel through, this hilltop implies that it is not an issue in the late universe where the Higgs is centered at the electroweak scale 246\,GeV. 

However, in the very early universe, it is anticipated that the universe was at such extremely high energies that the Higgs is in danger of fluctuation to the dangerous side of the potential.
Therefore, one should consider the serious possibility that the Higgs field could have fluctuated past $\hmax$ during an early era; of particular focus here was an early era of inflation. If this had occurred, the Higgs could have run down the potential towards infinity, and the universe would presumably undergo a catastrophic crunch.
Important earlier work in this area, includes Refs.~\cite{Espinosa.Riotto.1,Zurek.1,Zurek.2,Herranen:2015ima,Gross:2015bea,Ema:2016kpf,Enqvist:2016mqj,Kohri.1,Kohri.2,Jain:2019gsq,Jain:2019wxo}.

One may potentially avoid this problem by significantly lowering the scale of inflation or utilizing a large effective mass of the Higgs during inflation, which we shall discuss later. But there is still a possible disaster that can take place  after inflation during p/reheating. In particular, one anticipates a direct coupling between the Higgs and the inflaton, and this can cause (parametric or tachyonic) resonance during reheating that exponentially increases in the value of the Higgs. In fact a coupling between the inflaton and Higgs is required for efficient reheating of the SM as it is the only renormalizable way a (gauge singlet) inflaton can couple to the SM. Potential ways to handle this problem has been discussed in the literature, including Refs.~\cite{Figueroa:2015rqa,Joti:2017fwe,Ema:2016ehh,Markkanen:2018pdo,Saha:2016ozn,Ema:2017rkk,Gong:2017mwt,Hertzberg:2019prp}.

Altogether, constraints on inflaton-Higgs couplings are needed to determine if a simple model in which the SM is taken seriously to high scales is possible; this is the focus of this work. We will examine all dimension 4 couplings between the Higgs and the inflaton, and derive combined bounds on this parameter space from demanding stability both during and after inflation -- see ahead to Figure \ref{fig:combined} for a final result. We use this to place a bound on the reheating temperature and discuss possible consequences.

\section{Inflaton-Higgs Coupling}

In conventional models of inflation, the exponential expansion of the universe is driven by a heavy scalar field $\phi$ -- the inflaton. 
During inflation, the inflaton's potential was relatively constant with the field at very large values. As a result, the inflaton dominated the universe, and so the energy density of the early universe can be approximated as simply the sum of the inflaton's kinetic and potential energy
\begin{equation}
    \label{eq:energy density}
    \rho(t)=\frac{1}{2}\dot{\phi}^2+V_\phi(\phi).
\end{equation}
As the universe expanded, most fields fell down their potential towards zero. However, since the inflaton potential was roughly constant at this time, its field, and in turn, $\rho(t)$, would have also remained roughly constant. 

The Friedman equation
\begin{equation}\label{eq:fried}
    H^2(t)=\frac{\dot{a}^2(t)}{a^2(t)}=\frac{8\pi G}{3}\rho(t),
\end{equation}
where $H(t)$ is the Hubble parameter and $a(t)$ is the scale factor, says that if $\phi$ undergoes sufficient Hubble friction and moves slowly, $\rho(t)$ is nearly constant, so the solution for $a(t)$ is approximately exponential
    $a(t) \propto e^{Ht}$.

As the inflaton continues to roll, inflation will eventually end.
Our primary interest here is in the era immediately following the end of inflation, (although we shall also consider further bounds from the inflationary phase itself). During this subsequent era, the inflaton must have decayed into the SM in order to reheat the universe. Plausibly, this was dominated by the inflaton's decay into the Higgs as it is the only possible renormalizable channel (for a gauge singlet inflaton). All other particles in the SM must couple to the inflaton via higher order operators and therefore may be highly suppressed.

\subsection{Action and Inflaton Evolution}
\begin{figure}
    \centering
    \includegraphics[width=0.7\linewidth,angle=-1]{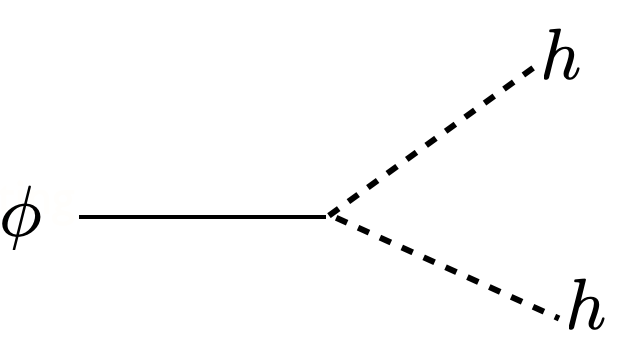}\\
        \includegraphics[width=0.73\linewidth,height=3cm,angle=-1.5]{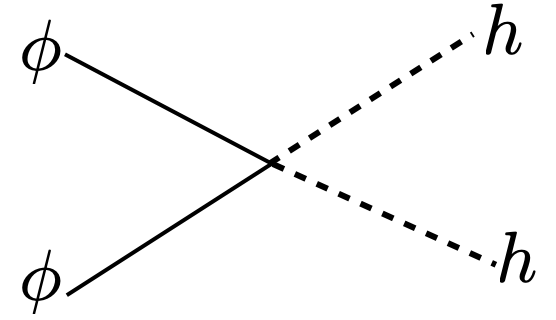}
    \caption{Upper: Feynman diagram for the inflaton decay into two Higgs particles (${\kappa\over2}\phi\, h^2$).
    Lower: Feynman diagram for inflaton annihilation into two Higgs particles ($\frac{\alpha}{2}\phi^2h^2$).}
    \label{fig:phi_h2}
\end{figure}

The action for gravity, the Higgs, and the inflaton is (units $\hbar=c=1$, signature $+---$) is given by
\begin{eqnarray}\label{eq:action}
    S&=&\int d^4x \sqrt{-g}\Big[\frac{-\mathcal{R}}{16\pi G_N}
    +\frac{1}{2}g^{\mu\nu}\partial_\mu h \partial_\nu h + \frac{1}{2}g^{\mu\nu}\partial_\mu\phi\partial_\nu\phi \nonumber\\
    &&\quad\quad\quad\quad\quad - V_h(h) -V_\phi(\phi)-\frac{\kappa}{2}\phi\, h^2  \Big]
\end{eqnarray}
The $-\kappa\,\phi\,h^2/2$ term provides perturbative inflaton decay into two Higgs particles as illustrated in the upper part of Figure \ref{fig:phi_h2}. Moreover, it is critical for understanding how the Higgs will grow during reheating, which can be resonant. The coefficient $\kappa$ (which has units of mass) specifies its strength. The only other allowed renormalizable interaction is $\sim -\alpha\,\phi^2h^2/2$ and will be discussed in Section \ref{Quartic} (see lower part of Figure \ref{fig:phi_h2}). Other terms in the action, including terms for all other SM particles and their interactions, may be small during reheating because they can only couple to the (gauge singlet) inflaton through higher dimension operators, and will not be included in our simplified analysis.

With the above action, the following is the Heisenberg equation of motion for the inflaton
\begin{equation}
    \label{eq:inflatoneom1}
    \ddot{\hat{\phi}}+3H(t)\dot{\hat{\phi}}+\frac{\partial V_\phi}{\partial\hat\phi}=0.
\end{equation}
To first approximation, the inflaton is a classical field with small quantum fluctuations
\begin{equation} \label{phiapprox}
    \hat{\phi}(t) = \phi(t) + \delta\hat{\phi}(t).
\end{equation}
Since the quantum fluctuations relative to the classical background are very small in standard inflationary models, the inflaton can be approximated as simply its classical, background field $\phi(t)$ when we study the impact on the Higgs.

During reheating, the inflaton's potential can be approximated by a mass term around its minimum
\begin{equation}
    \label{eq:inflaton potential}
    V_\phi(\phi)=\frac{1}{2}m_\phi^2 \phi^2
\end{equation}
Using this potential in the (classical version of) Eq.~\ref{eq:inflatoneom1} gives
\begin{equation}
    \label{eq:inflatoneom}
    \ddot{\phi}+3H(t)\dot{\phi}+m_\phi^2 \phi=0.
\end{equation}
So the equation of motion is that of a damped harmonic oscillator with Hubble acting as a friction term.

\subsection{Higgs Evolution Equations}

The Heisenberg equation of motion for the Higgs is
\begin{equation}
    \label{eq:higgseom}
    \ddot{\hat{h}}+3H(t)\dot{\hat{h}}-\frac{\nabla^2\hat{h}}{a^2(t)}+\frac{\partial V_h}{\partial\hat{h}}+\kappa\,\phi(t)\hat{h}=0.
\end{equation}

For simplicity, the partial derivative of the Higgs potential can be dropped from the equation of motion. It turns out that this term is around five orders of magnitude smaller than $\kappa\phi\hat{h}$ when we expand around the Higgs vacuum. It therefore has a negligible effect during the first stages of evolution. Of course, once the inflaton has completely decayed into the Higgs, the $\kappa\,\phi\,\hat{h}$ term goes away and the shape of the Higgs potential becomes critical. If the Higgs is on the other side of the hilltop at that point, then it would run down the potential and cause a disaster. Therefore, while the Higgs potential can be neglected during the first stages of reheating, it will be critical in our understanding of the final analysis. Therefore we shall incorporate its effects qualitatively by noting that this problem can occur. Ultimately, a more precise analysis would involve the inclusion of this term throughout the entire analysis. This renders the equations nonlinear, with a $\sim\lambda\, \hat{h}^3$ term. This could be handled with full lattice simulations. However, this is beyond the scope of the current work. Our results are therefore approximate and can be improved upon with more detailed simulations. However, the qualitative and semi-quantitative results found here are anticipated to provide a reasonable estimate.

In this absence of the regular Higgs potential term, the resulting equation is linear. So it is useful to take a Fourier transform and the equation of motion is then given in terms of the wavenumber, $k$, as
\begin{equation}
    \label{eq:higgseom2}
    \ddot{\hat{\tilde{h}}}_k+3H(t)\dot{\hat{\tilde{h}}}_k+\frac{k^2}{a^2(t)}\hat{\tilde{h}}_k+\kappa\,\phi(t)\hat{\tilde{h}}_k=0.
\end{equation}
Since the equation of motion is now linear with respect to the Higgs, and the background metric and inflaton is approximated as carrying spatial translation invariance, the Higgs field can be solved in terms of mode functions $v_k(t)$, as follows
\begin{equation}
    \label{eq:higgsdecomp}
    \hat{h}(\vec{x},t)=\int \frac{d^3k}{(2\pi)^3} \left[(v_k(t)\hat{a}_k + v_{-k}^*(t)\hat{a}_{-k}^\dagger)e^{i\vec{k}\cdot\vec{x}}\right].
\end{equation}
Since the annihilation and creation operators are constant in time, they can be factored out, and thus the mode functions must satisfy the same equation of motion
\begin{equation}
    \label{eq:modeeom}
    \ddot{v}_k+3H(t)\dot{v}_k+\frac{k^2}{a^2(t)}v_k+\kappa\,\phi(t)v_k=0.
\end{equation}
Thus need to solve this set of ordinary differential equations (ODEs) in order to solve the theory at the linear level. 

In the free theory, where the Higgs and inflaton are not coupled and $\kappa = 0$, the solution to this equation of motion is
\begin{equation}
    \label{vkinit}
    v_k(t)=\frac{1}{a_i\sqrt{2k}}\textrm{exp}\left(\frac{-ikt}{a_i}\right)
\end{equation}
where the particular solution is chosen to maintain the Higgs' commutation relations. In the coupled theory, this solution is an initial condition which the Higgs evolves away from as the inflaton evolves. 

We are interested in computing the evolution, as the coupling to the inflaton may cause a radical growth in its variance. This will imply the Higgs would have a significant probability of being on the dangerous side of the effective potential $V_h$.
The variance is readily calculated from Eq.~\ref{eq:higgsdecomp} to be
\begin{equation}
    \label{eq:variance}
    \langle \hat{h}^2 \rangle = \int \frac{d^3k}{(2\pi)^3} \lvert v_k(t) \rvert^2.
\end{equation}

In the initial vacuum state and in absence of coupling, the variance in the Higgs field can be calculated by inserting Eq.~\ref{vkinit} into Eq.~\ref{eq:variance}
\begin{equation}
    \langle h^2 \rangle_\tf = \int \frac{d^3k}{(2\pi)^3}\frac{1}{2a_i^2 k}
\end{equation}
or, after integrating out the angular components,
\begin{equation}
    \langle h^2 \rangle_\tf = \int \frac{dk}{4a_i^2\pi^2} k.
\end{equation}
While this integral is formally UV divergent, we will be only interested in a smoothed field with a cutoff set by a bubble size (to be discussed in Section \ref{Bubble}), so in fact the relevant fluctuations are finite (the two-point correlation function is finite on the scales of interest). Introducing non-zero values of $\kappa$ corresponds to solutions that evolve away from Eq.~\ref{vkinit}. By dividing out the variance of the interacting theory by the variance of the free theory, the {\em growth} in variance can be determined. 

\section{Simplified Analysis }
In the next Section, we shall solve these equations numerically. However, for now it is useful to gain intuition and develop approximate semi-analytical results.

The solution to Eq.~\ref{eq:modeeom} is different for each value of $k$. We will choose a discrete values for $k$ and then sum the results. Eq.~\ref{eq:modeeom} can first be approximated by treating the time dependent prefactors as varying adiabatically slowly. This means it becomes a type of  Mathieu equation. Solving this gives a better understanding of the best range of $k$ values to sum over. The Mathieu equation has the canonical form
\begin{equation}
    \label{eq:mathieu}
    {d^2y\over d\tau^2}(\tau) + \left(A-2q \cos(2\tau)\right)y(\tau)=0
\end{equation}
Its solutions are of the form
\begin{equation}
    \label{eq:mathieu soln}
    y(\tau)=e^{\mu\tau}f_1(\tau)+e^{-\mu\tau}f_2(\tau)
\end{equation}
where $f_{1,2}(\tau)$ are periodic. If the so-called Floquet exponent $\mu$ is real the growth is exponential, otherwise the growth is absent. 

\begin{figure}[t]
    \centering
        \includegraphics[width=\linewidth,height=6.4cm]{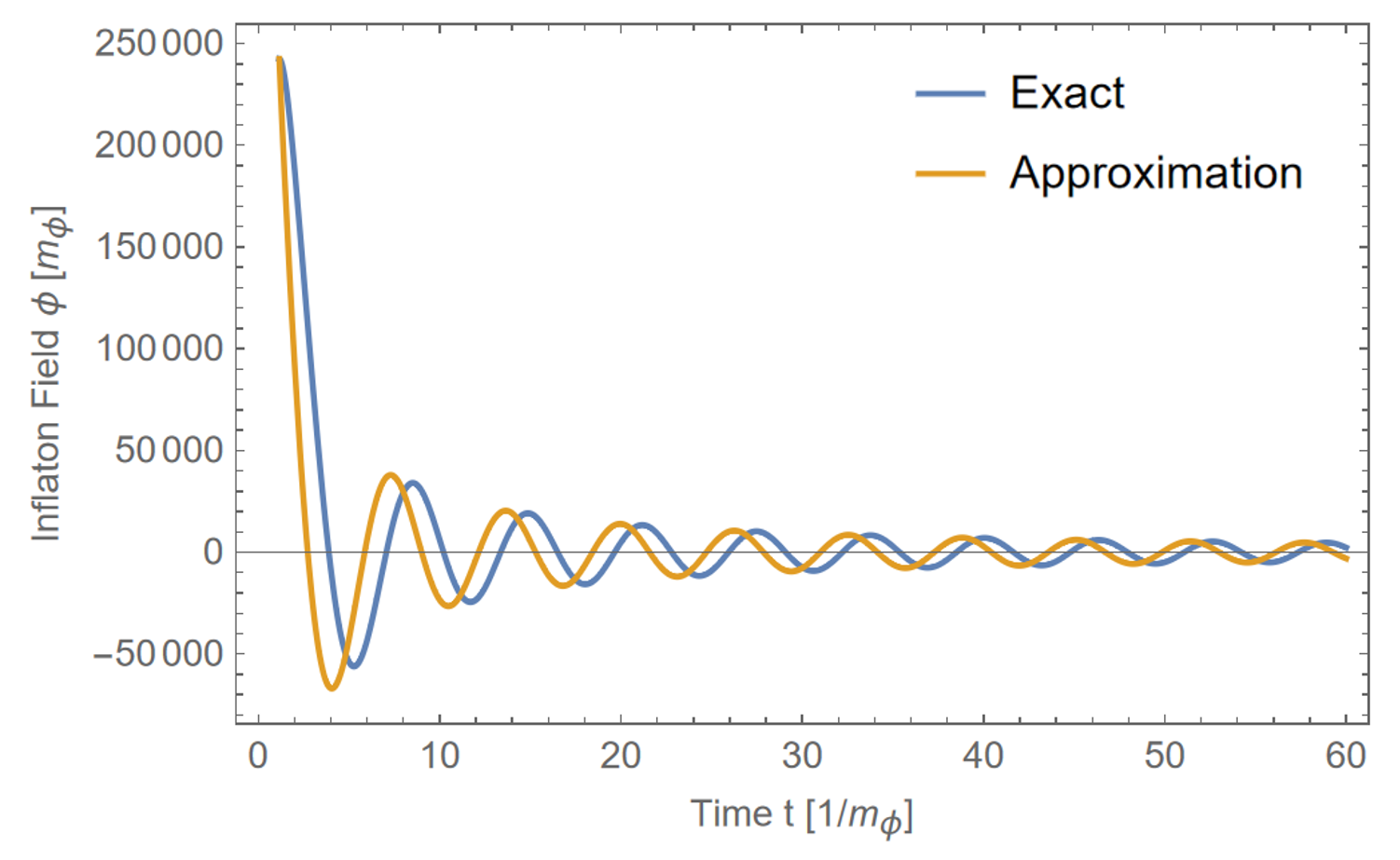}
    \includegraphics[width=0.96\linewidth,height=6.2cm]{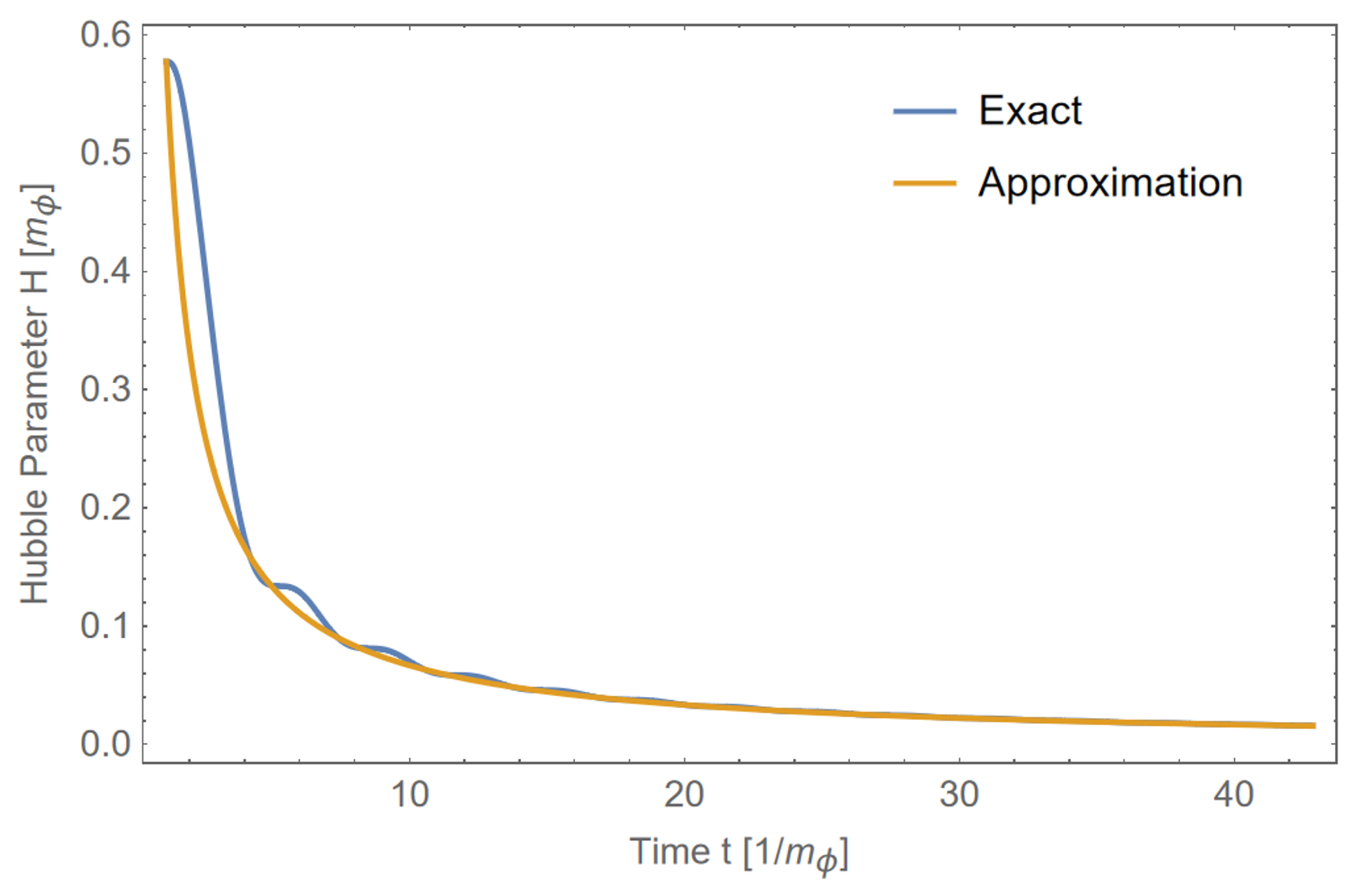}
    \caption{
   Upper: Comparison of exact and approximate solutions of the inflaton field.
    Lower: Comparison of exact and approximate solutions of the Hubble parameter.}
    \label{fig:hubble approx}
\end{figure}

The Mathieu equation requires a harmonic driving term which can come from the inflaton field. As in Eq.~\ref{eq:inflaton potential}, the inflaton potential resembles that of a harmonic oscillator during reheating, and so the field can be approximated as
\begin{equation}
    \label{eq:inflaton approx}
    \phi(t) \approx \phiamp(t)\cos(m_\phi t).
\end{equation}
The inflaton's equation of motion \ref{eq:inflatoneom}, includes a Hubble friction term. This can be incorporated in this approximation by allowing the envelope $\phiamp$ to decrease with time as the universe expands. In particular, since reheating is approximately a matter dominated era (as the inflaton's oscillations lead to the pressure averaging to zero), the Hubble parameter during this time can be approximated as
\begin{equation}
    \label{eq:hsimple}
    H(t)\approx \frac{2}{3t},
\end{equation}
so that
\begin{equation}
    \label{eq:asimple}
    a(t) \approx a(t_i) \left(\frac{t}{t_i}\right)^{2/3}.
\end{equation}
This approximation is not precise right at the start of reheating, the end of inflation, but becomes more accurate over time as the universe transitions from the accelerating phase to an effective matter dominated phase. Under these approximations $\phiamp(t)$ decreases as
\begin{equation}
    \label{phiamp}
    \phiamp(t) = \phi_i\left(\frac{a(t_i)}{a(t)}\right)^{3/2}.
\end{equation}
To ensure that these are reasonable approximations, they are compared to numerical solutions for $H(t)$ and $\phi(t)$ from solving Eqs.~\ref{eq:energy density}, \ref{eq:fried}, and \ref{eq:inflatoneom}. This is shown in Figure \ref{fig:hubble approx}. As expected, these approximations become increasingly accurate at later times (the late time off-set in $\phi$ is just a phase shift).

To generate these and other plots, we take the inflaton mass to be $m_\phi \approx 1.4 \times 10^{13}$ GeV (a value expected in chaotic inflation \cite{Linde:2007fr}). Calculations are done in units of $m_\phi$, such that the reduced Planck mass is $\Mpl = 1/\sqrt{8\pi G}=2.4 \times 10^{18}\,\mbox{GeV} \approx 1.7\times 10^5\,m_\phi$, and $\phi_i = \sqrt{2}\,\Mpl$ is the field value at the end of inflation under the quadratic potential approximation.

In addition to approximating the inflaton field as a classically oscillating function, the expansion terms that depend on time must be removed from Eq.~\ref{eq:modeeom} in order to approximate it as a Mathieu equation. The adiabtic approximation involves replacing in the inflaton field by decreasing $\phiamp$ with time according to Eq.~\ref{phiamp} so as to capture its effects. Similarly, $a(t)$ can be removed from the equation of motion with the following substitution
\begin{equation}
    \kphys(t) = \frac{k}{a(t)}.
\end{equation}
Here, $k_{phys}(t)$ is the physical wavenumber at a particular point in time and decreases as the universe expands. By allowing $k_{phys}(t)$ and $\phiamp(t)$ to decrease with time, expansion will be approximately captured in the solution, and so the Hubble term can be dropped in this first analysis.

\begin{figure}[t]
    \centering
    \includegraphics[width=\linewidth]{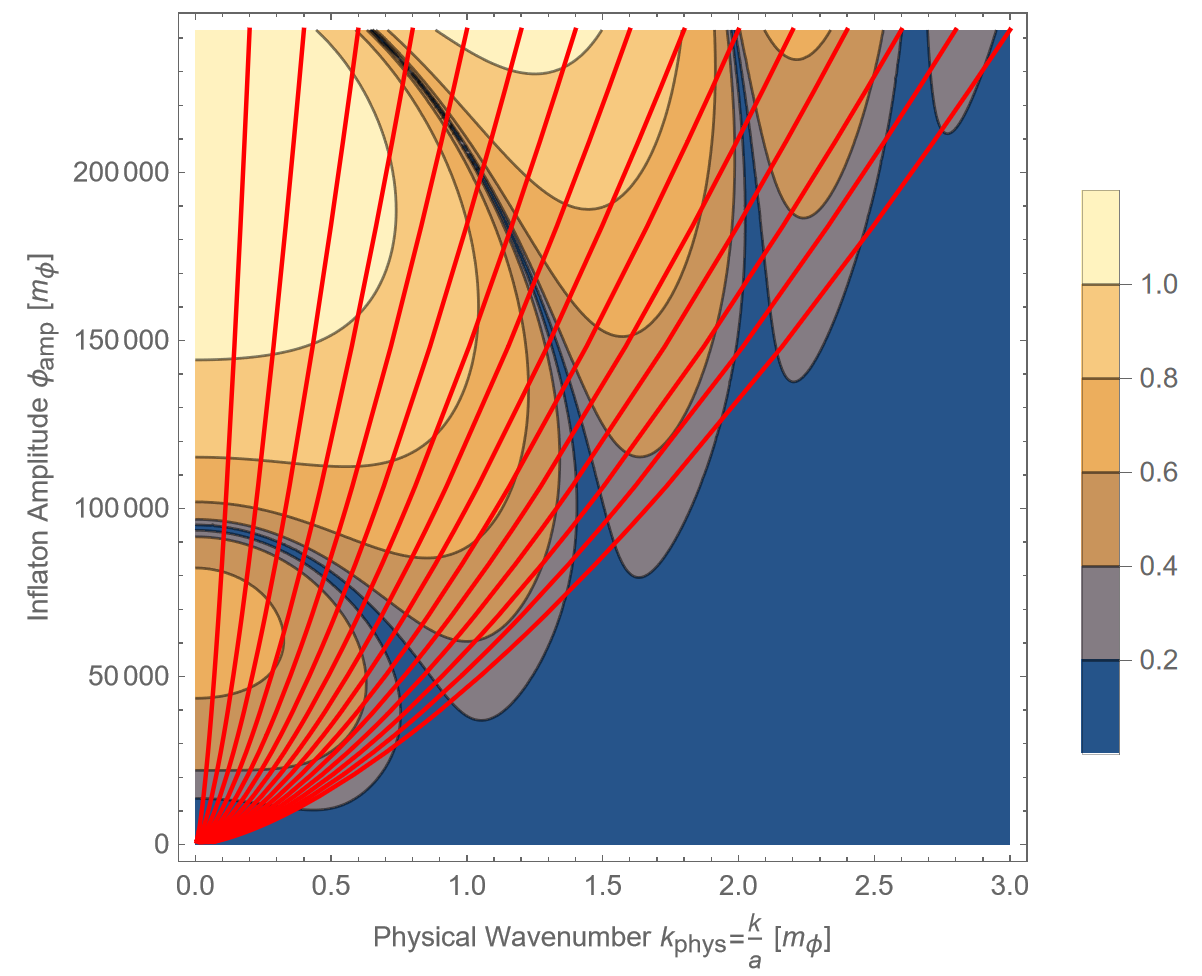}
    \caption{Floquet exponent for Higgs resonance due to cubic coupling decay for $\kappa = 4 \times 10^{-5}\,m_\phi$. Brighter yellow regions represent regions of greater resonance and a larger characteristic exponent as specified by the color legend. The red lines represent how a particular wavenumber $k$ evolves as $\kphys(t)$ and $\phiamp(t)$ decrease with the expansion of the universe.}
    \label{fig:kappa_mathieu}
\end{figure}

Substituting these approximations into Eq.~\ref{eq:modeeom} gives 
\begin{equation}
    \label{eq:mode mathieu}
    \ddot{v}_k(t) + \left(\kphys^2+\kappa\phiamp \cos(m_\phi t)\right) v_k(t) = 0.
\end{equation}
Comparing Eq.~\ref{eq:mode mathieu} to Eq.~\ref{eq:mathieu}, the following change of variables must be made to reproduce a Mathieu equation 
\beq
\label{change of variables}
    \tau = \frac{m_\phi t}{2}, \quad A = \frac{4\kphys^2}{m_\phi^2}, \quad q = \frac{2\,\kappa\,\phiamp}{m_\phi^2}.
\eeq
The resulting Floquet exponents $\mu$ are plotted as contours in Figure \ref{fig:kappa_mathieu} as a function of $\kphys(t)$ and $\phiamp(t)$ for $\kappa = 4 \times 10^{-5}$ $m_\phi$. The red lines in Figure \ref{fig:kappa_mathieu} represent a discrete set of values of comoving wavenumber $k$, which redshift to the left according to $k_{phys}(t)\propto 1/a(t)$ and downwards according to $\phiamp(t)\propto 1/a(t)^{3/2}$ as the universe expands. From this approximation and further numerical analysis, the range $0.2 \leq k/m_\phi \leq 3$ and resolution $\Delta k/m_\phi = 0.2$, was found to be sufficient as all these lines pass through the yellow bands of significant resonance. 

\section{Full Numerical Analysis}
\begin{figure}[t]
    \centering
    \includegraphics[width=\linewidth,height=6.4cm]{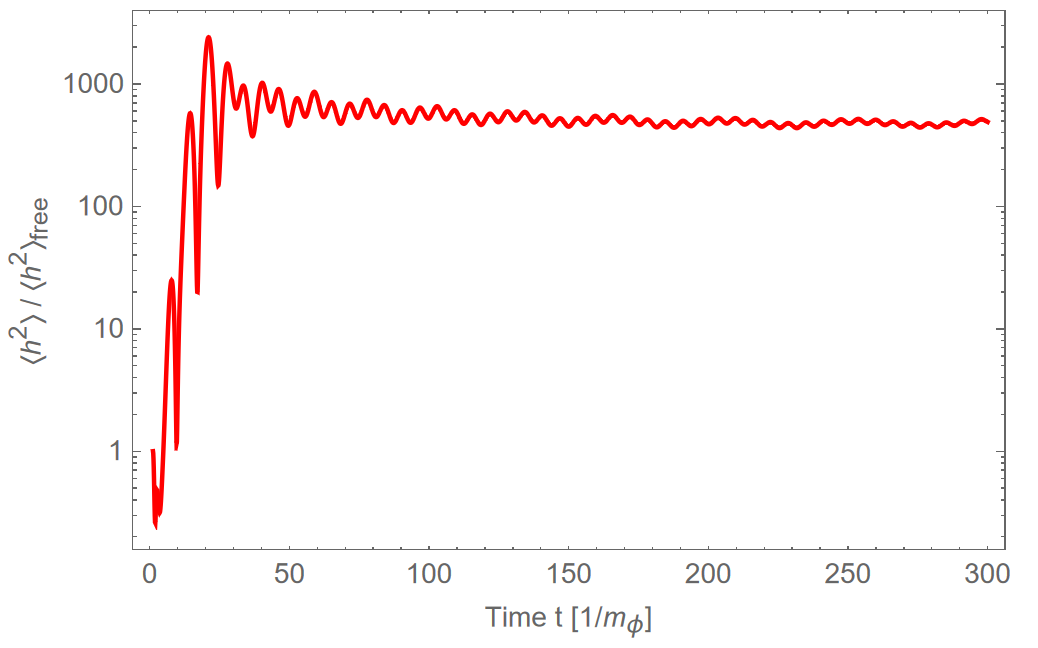}
        \includegraphics[width=\linewidth,height=6.4cm]{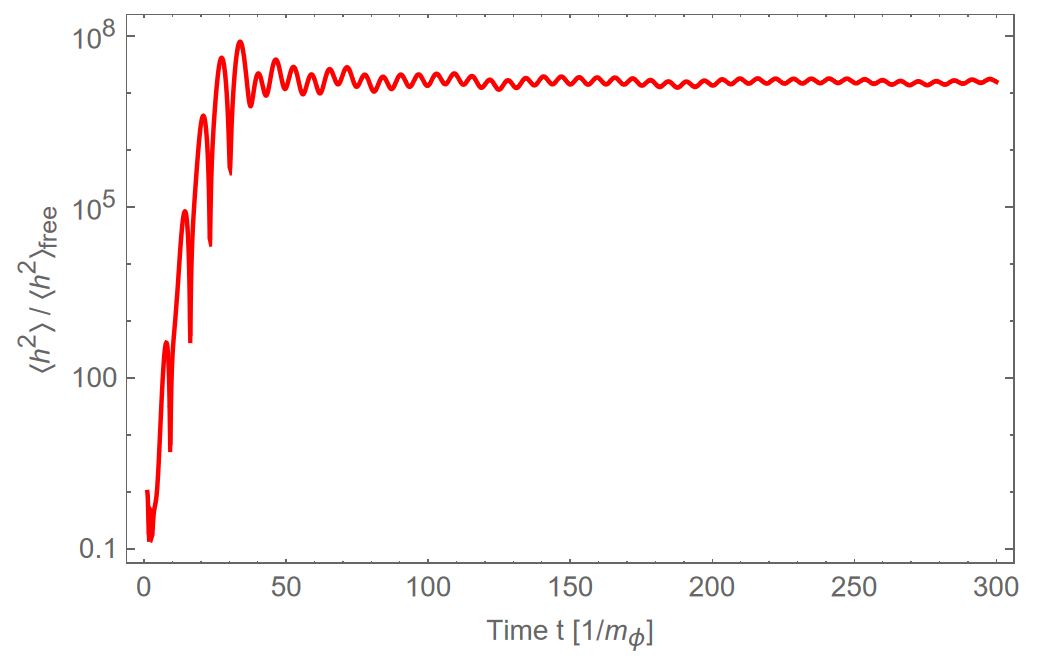}
    \caption{Upper: Growth in Higgs variance $\langle h^2 \rangle / \langle h^2\rangle_\tf$ with respect to time for $\kappa=2\times 10^{-5}$ $m_\phi$.
    Lower: Growth in Higgs variance $\langle h^2 \rangle / \langle h^2\rangle_\tf$ with respect to time for $\kappa=4\times 10^{-5}$ $m_\phi$.
 }
    \label{fig:kappa2}
\end{figure}

\begin{figure}[t]
    \centering
                \includegraphics[width=\linewidth]{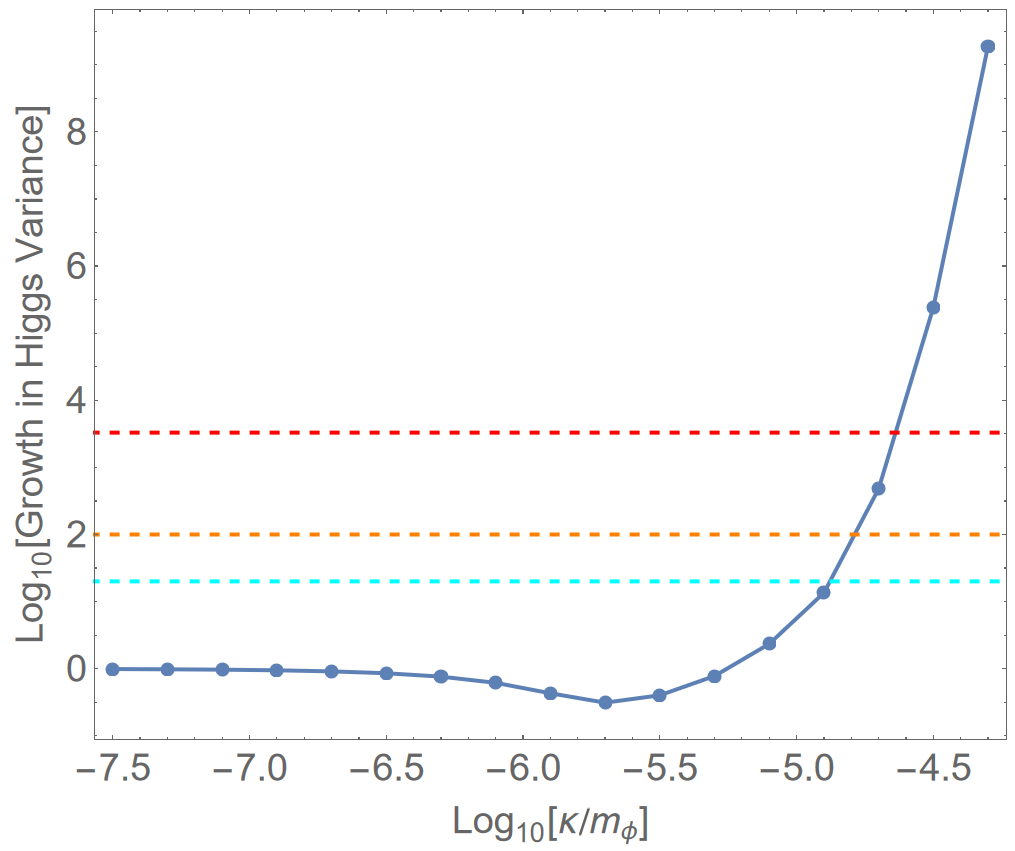}
\caption{Late time growth in the Higgs variance $\langle h^2 \rangle / \langle h^2\rangle_\tf$ as a function of $\kappa/m_\phi$.
    Dotted lines represent where the Higgs field grows by a factor of 20 (cyan), 100 (orange), and 3300 (red).}
    \label{fig:kappaLate}
\end{figure}

With a range of $k$ values chosen, $v_k(t)$ can be numerically calculated according to Eq.~\ref{eq:modeeom} and the Higgs variance can be calculated according to Eq.~\ref{eq:variance}. The growth in the Higgs field is measured as the ratio between the variance in the coupled theory and the variance in the free theory
\beq
\mbox{growth}= {\langle h^2\rangle\over\langle h^2\rangle_\tf}.
\eeq
For two values of $\kappa$, this growth is shown in the upper and lower panels of Figure \ref{fig:kappa2}.
These plots illustrate how sensitive the Higgs field is to a small change in $\kappa$ as a factor of two causes a difference in growth of $10^5$. To determine if the Higgs field will grow past $h_{max}$, only the endpoint of the variance plots need to be considered. Figure \ref{fig:kappaLate} shows the growth of the Higgs field at late times as a function of $\kappa$. There appears to be a dip in Higgs growth near $\kappa \approx 10^{-5.7}\,m_\phi$ where the Higgs field apparently decreases at late times. This is possible is because the inflaton is oscillating, and so resonance may just coincidentally occur when the inflaton is at a minimum for this  value of $\kappa$. But more typically, larger $\kappa$ tends to lead to much larger growth. In the following section, the amount of growth required for the Higgs to go over its hilltop will be discussed.

\section{Creating a Bubble}\label{Bubble}
The probability distribution of any ground state in quantum mechanics with a quadratic potential is Gaussian. The Higgs field's probability distribution can therefore be approximated as a Gaussian, since its potential is primarily quadratic with a quartic term that only contributes a small correction. So, the probability that the field would grow past $\hmax$ where the potential $V_h(h)$ peaks is approximately
\begin{equation}
    \label{gauss prob}
    P(h\geq \hmax) \propto \exp\left(-\frac{\hmax^2}{2\langle h^2 \rangle}\right).
\end{equation}
where we use the fact that we will be typically in the tail of the distribution, so an integral past the peak provides a small correction.

To compute this value, it is important to first note that an individual point in the Higgs field cannot grow independently of the points around it due to field gradients. Rather, a type of bubble will form with a characteristic size, $\Rbub$, and grow. 
The size of the bubble depends on its height. Numerically, this can be explained by energy conservation. The energy density of the Higgs field is given by
\begin{equation}
    \label{higgs energy density}
    \rho_h = \frac{1}{2}\dot{h}^2 +\frac{1}{2}(\nabla h)^2 + V_h(h)
\end{equation}
where $V_h(h)$ is defined by Eq.~\ref{eq:higgspot}. As the field fluctuates up the potential, the field around it will tend to grow as well so that the gradient and time derivatives balance this increase in the potential and the energy density does not change too quickly. Given that the field's gradient is approximately equal to its time derivative (for such relativistic bubbles), this balancing act requires that $\frac{1}{2}(\nabla h)^2 \sim V_h(h)$. Noting that the slope of the bubble is approximately its height divided by its characteristic radius, the gradient can be approximated as
\begin{equation}
    (\nabla h)^2 \sim \frac{\hbub^2}{\Rbub^2}
\end{equation}
Therefore, balancing the gradient with the potential means the height of a bubble is approximately proportional to the inverse of the bubble's size
\beq
    \frac{1}{2}\frac{h^2}{\Rbub^2} \sim \frac{|\lambda|}{4}\hbub^4,
        \label{hbubble}
\eeq
    which implies
    \beq
    \hbub \sim \frac{\sqrt{2}}{\Rbub\sqrt{|\lambda|}}.
\eeq

Furthermore, the variance $\langle h^2 \rangle$ is also proportional to $1/\Rbub$. This is because when calculating the variance for a bubble according to Eq.~\ref{eq:variance}, the integral only needs to be computed up to wavelengths of about the size of the bubble. Shorter wavelengths would not affect the overall shape of the bubble and can therefore be neglected. For the vacuum, this gives a variance of 
\beq
    \langle h^2 \rangle_\tf \sim \frac{1}{4\pi^2}\int_0^{\frac{2\pi}{\Rbub}} dk \, k = \frac{1}{2\Rbub^2}.
    \label{vacuum fluctuations}
\eeq
So, it is clear that the size of the bubble will drop out from the final probability calculations. 

A more precise analysis has determined that the ratio of field height to vacuum fluctuations is indeed of this order, but with a corrected $\mathcal{O}(1)$ prefactor \cite{Linde:1991sk}
\begin{equation} \label{prob}
    \frac{\hbub^2}{2\langle h^2 \rangle_\tf} = \frac{8\pi^2}{3|\lambda|}.
\end{equation}

\begin{figure}[t]
    \centering
    \includegraphics[width=\linewidth,height=6.2cm]{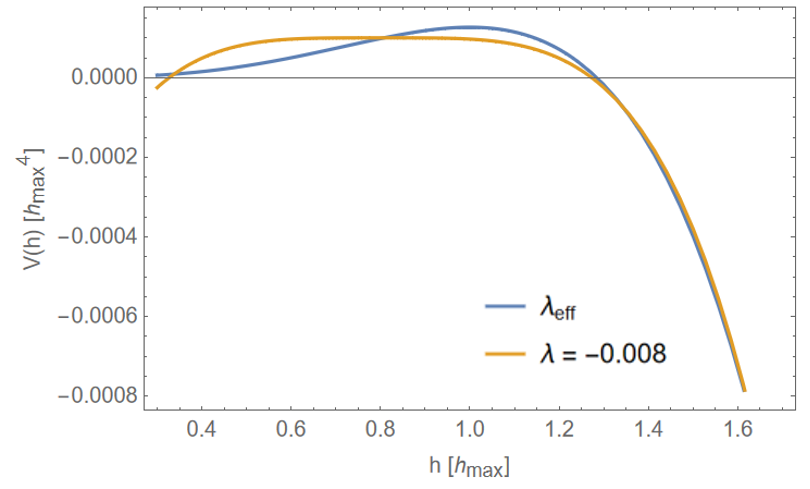}
    \caption{Comparison of $V_h(h) = \lambda(h)h^4/4$ for the fitting function $\lambda_\text{eff}(h)$ versus using a simple constant $\lambda = -0.008$. The latter function requires some translations in field value in order to show this approximate matching, but it is only the curvature determined by $\lambda$ that is of direct importance here.}
    \label{fig:lambda}
\end{figure}

Generally, $\lambda$ is a function of the Higgs field $h$, due to the renormalization group flow of the effective potential. It is convenient to have an approximate fitting function for it. A useful form is \cite{Espinosa.Riotto.1}
\begin{equation}
    \label{eq:lambda}
    \lambda_\text{eff} \approx -\frac{0.16}{(4\pi)^2}\ln\!\left(\frac{h^2}{\hmax^2\sqrt{e}}\right).
\end{equation}
For simpler calculations, $\lambda$ can be approximated as a constant $\lambda \approx -0.008$. This gives a potential that is somewhat accurate in the regime $V_h(h) < 0$, as seen in Figure \ref{fig:lambda}, which is the region of interest due to the turn-over in the Higgs potential.

In the vacuum, Eq.~\ref{prob} can therefore be directly plugged into Eq.~\ref{gauss prob} with 
\beq
\hbub\to\hmax
\eeq 
and $\lambda = -0.008$ to find the probability of the Higgs going over its hilltop today, when the inflaton coupling has no influence. The result is that $P \sim e^{-3300}$ which recovers the well known result that the Higgs in the ordinary vacuum today is very stable.

\section{Limits on Cubic Coupling}
To find the probability of the Higgs reaching $\hmax$ after resonance after inflation, the vacuum variance in Eq.~\ref{prob} (with $\hbub\to\hmax$) needs to be rescaled by the increased variance after resonance, which we earlier computed. For a reasonable approximation, the probability can therefore be given as 
\beq
    \label{gauss prob res}
    P_\text{res}(h\ge\hmax) \propto 
    \exp\left[-\frac{8\pi^2}{2|\lambda|} \times \left(\frac{\langle h^2 \rangle}{\langle h^2\rangle_\tf}\right)^{-1}\right].
\eeq
A very generous limit would be to demand that
\begin{equation} \label{boldest limit}
    P_\text{res} < e^{-1}
\end{equation}
which would put the Higgs field at $\sim40\%$ chance of growing past $\hmax$ and becoming unstable. Solving for the growth in variance for this limit gives
\begin{equation}
    \frac{\langle h^2 \rangle}{\langle h^2\rangle_\tf} < 3300.
\end{equation}
Meanwhile, the most conservative limits would have no more than one Hubble patch go over the hilltop. This is a more reasonable requirement, because even if one patch goes over it could expand to cause the rest of the universe into a crunch. Since there were about $\sim e^{3N_e}$ Hubble patches, where $N_e \approx 55$, this would correspond to
\beq
    P_\text{res} < e^{-3N_e} \\ \label{conservative}
    \eeq
    which leads to
    \beq
    \frac{\langle h^2 \rangle}{\langle h^2\rangle_\tf} < 20.
\eeq
If one Hubble patch were to fall down the Higgs potential, however, the patches around it could pull it back to the safe side. Therefore, a more reasonable estimate may allow for a little more growth than Eq.~\ref{conservative}
\begin{equation} \label{reasonable}
    \frac{\langle h^2 \rangle}{\langle h^2\rangle_\tf} < 100.
\end{equation}
Each of these limits is indicated in Figure \ref{fig:kappaLate}, where the horizontal dotted lines represent where the Higgs field growth is 20 (cyan), 100 (orange), and 3300 (red).

To satisfy the middle requirement \ref{reasonable}, we find the bound on the cubic coupling $\kappa$ is
\begin{equation} \label{kappalimit}
    \kappa < 1.6\times 10^{-5} m_\phi\sim 2.2 \times 10^8\,\mbox{GeV}.
\end{equation}
(This result can be compared to Ref.~\cite{Enqvist:2016mqj} that also placed a bound on this cubic coupling. Related work also includes Refs.~\cite{Espinosa.Riotto.1,Zurek.1,Zurek.2,Herranen:2015ima,Gross:2015bea,Ema:2016kpf,Kohri.1,Kohri.2,Jain:2019gsq,Jain:2019wxo,Figueroa:2015rqa,Joti:2017fwe,Ema:2016ehh,Markkanen:2018pdo,Saha:2016ozn,Ema:2017rkk,Gong:2017mwt,Hertzberg:2019prp}.)
Here we are assuming $\kappa>0$; but the case of negative $\kappa$ is almost identical as the inflaton oscillates, so the sign is relatively unimportant. So the complete bound is $|\kappa| < 1.6\times10^{-5} m_\phi\sim2.2 \times 10^8\,\mbox{GeV}$. However, in Section \ref{DuringInflation}, the positive $\kappa$ case will be the focus.

\section{Bound on Reheat Temperature}
Limits on inflaton-Higgs couplings can also put constraints on the reheat temperature of the universe. First, the inflaton decay rate can be calculated under the reasonable assumption that a gauge-singlet inflaton primarily decays into the Higgs. This is because it is only to the Higgs that one can form renormalizable couplings. While other interactions are dimension 5 and above, which are plausibly suppressed by a very high scale, such as $\Mpl$. At late times, after any resonance has occurred, perturbative tree-level decay into the Higgs from the cubic coupling can be most important
\begin{equation}
    \label{decay}
    \Gamma(\phi\rightarrow hh) = \frac{g_h}{32\pi}\frac{\kappa^2}{m_\phi}
\end{equation}
where $g_h = 4$ is the number of components in the Higgs. Using the bound on $\kappa$ in Eq.~\ref{kappalimit} and assuming an inflaton mass of $\sim1.4\times 10^{13}$ GeV, this corresponds to a bound on the decay rate of $\Gamma(\phi\rightarrow hh) \lesssim 2.2\times 10^{26}\,\mbox{sec}^{-1}$. 

As the inflaton decays, the large amount of energy stored in its mass is converted into kinetic energy in the significantly lighter Higgs and then the rest of the SM, producing a bath of relativistic particles. These particles will rapidly thermalize at time $t \sim 1/\Gamma$. Since Hubble at this time is around $H \sim 1/t$ as the universe becomes matter dominated during reheating, and so it is true that $H \sim \Gamma$ at this time. Given an approximate value for $H$, the energy density of the thermal bath can be calculated from the Friedman equation (eq. \ref{eq:fried}). The temperature of this thermalized sea of radiation can then be calculated from its energy density
\begin{equation}
    \rho = \frac{\pi^2}{30} \,g\, T^4
\end{equation}
where $g =106.75$ represents the degrees of freedom in the SM (if other new degrees of freedom are present, we assume it does not increase the total $g$ too much). These calculations give a reheat temperature of 
\begin{equation}
    \Tr \approx 0.5\sqrt{\Gamma M_{Pl}}.
\end{equation}

Therefore, the calculated constraints on $\kappa$ from Eq.~\ref{kappalimit} gives a reheat temperature bound of
\begin{equation}
    \Tr \lesssim 9.2 \times 10^9\,\mbox{GeV}.
\end{equation} 
where the uncertainty arises from the uncertainty in inflaton mass. (This can be compared to the work in Ref.~\cite{Enqvist:2016mqj} that also placed a bound on the reheat temperature.) Interestingly, this temperature is well below the temperature required in GUT scale models of baryogenesis that may help to explain the matter to anti-matter asymmetry; so this is potentially quite important.

\section{Including Quartic Coupling}\label{Quartic}

While the resonance due to the $-\kappa\,\phi\,h^2/2$ term in Eq.~\ref{eq:action} may dominate the growth in the Higgs field, an inflaton annihilation term $-\alpha\,\phi^2h^2/2$ in the action can also be considered
\begin{eqnarray}    \label{alpha action}
    S&=&\int d^4x \sqrt{-g}\Big[\frac{-\mathcal{R}}{16\pi G_N}
    +\frac{1}{2}g^{\mu\nu}\partial_\mu h \partial_\nu h + \frac{1}{2}g^{\mu\nu}\partial_\mu\phi\partial_\nu\phi \nonumber\\
    &&\quad\quad\quad\quad\quad - V_h(h) -V_\phi(\phi)-\frac{\kappa}{2}\phi\, h^2 -\frac{\alpha}{2}\phi^2 h^2 \Big]
\end{eqnarray}
The Feynman diagram for this annihilation is given in the lower part of Figure \ref{fig:phi_h2}.
This changes the equation of motion for the mode functions to
\begin{equation}
    \label{eq:alphaeom}
    \ddot{v}_k(t)+3H(t)\dot{v}_k+\frac{k^2}{a^2(t)}v_k+\kappa\,\phi(t)v_k+\alpha\,\phi^2(t)v_k=0.
\end{equation}
Following the same steps as before, the Mathieu approximations can be used again to analyze the effects of varying $\alpha$ by first setting $\kappa = 0$ and noticing that
\begin{figure}[t]
    \centering
    \includegraphics[width=\linewidth]{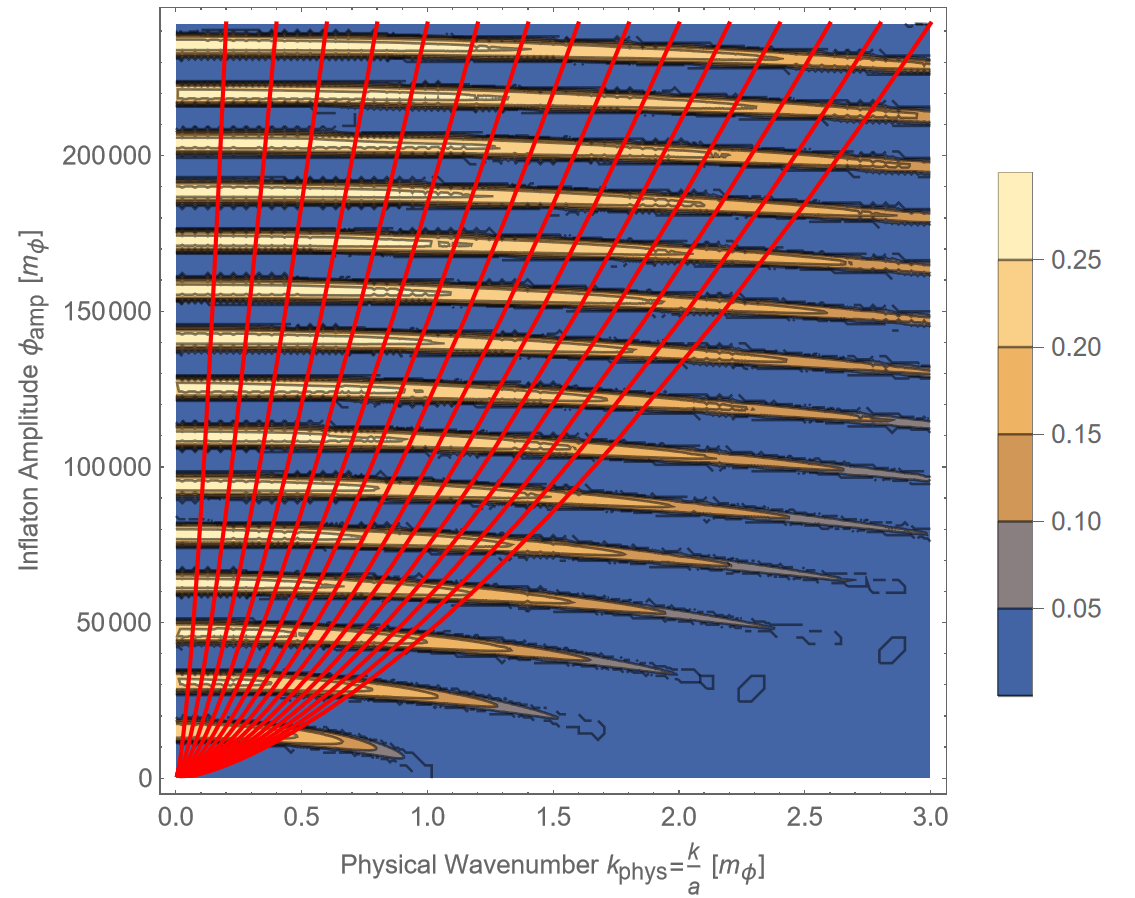}
    \caption{Floquet exponent for Higgs resonance due to quartic coupling for $\alpha = 10^{-8}$ (and $\kappa=0$). Brighter yellow regions represent regions of greater resonance and a larger characteristic exponent as specified by the color legend. The red lines represent how a particular wavenumber $k$ evolves as $\kphys(t)$ and $\phiamp(t)$ decrease with the expansion of the universe.}
    \label{fig:alphamathieu}
\end{figure}
\begin{align}
    \begin{split}
        \phi(t)^2 &\approx \phiamp(t)^2\cos^2(m_\phi t) 
        = \frac{\phiamp(t)^2}{2}(1 + \cos(2m_\phi t)).
    \end{split}
\end{align}
Thus, the change of variables for the pure $\alpha$ theory analogous to Eq.~\ref{change of variables} are
\begin{equation}
    \tau = m_\phi t, \quad A = \frac{\kphys^2}{m_\phi^2} + \frac{\alpha}{2m_\phi^2}\phiamp^2, \quad q = \frac{\alpha}{4m_\phi^2}\phiamp^2.
\end{equation}
The Floquet exponent for $\alpha = 10^{-8}$ (with $\kappa=0$) with respect to $\phiamp$ and $\kphys$ is shown in Figure \ref{fig:alphamathieu}.

\begin{figure}[t]
    \centering
    \includegraphics[width=\linewidth]{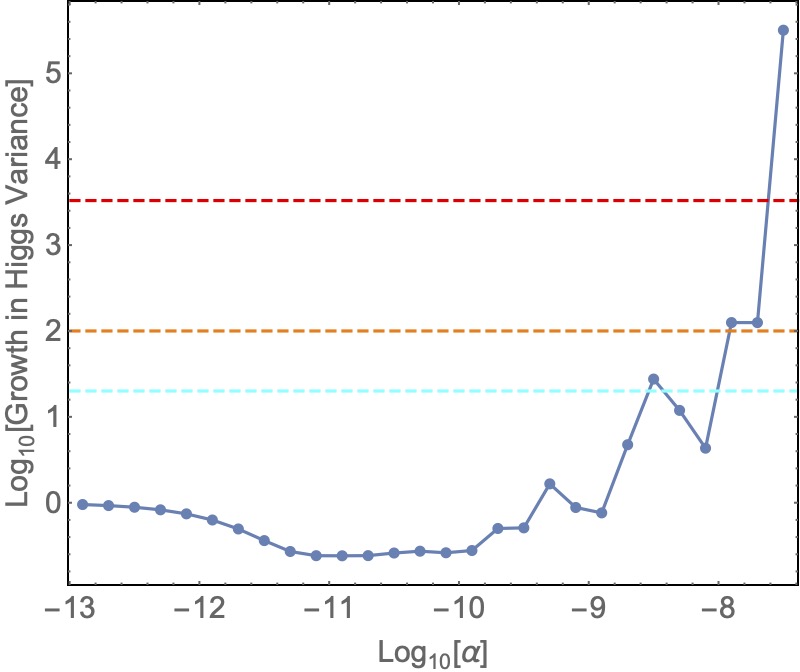}
    \caption{Late time growth in Higgs variance $\langle h^2 \rangle/ \langle h^2\rangle_\tf$ as a function of $\alpha$ (with $\kappa = 0$). Dotted lines represent where the Higgs field grows by a factor of 20 (cyan), 100 (orange), and 3300 (red).}
    \label{fig:growthalpha}
\end{figure}

Figure \ref{fig:growthalpha} shows the numerically calculated growth for a pure $\alpha$ theory as well as the previously discussed constraints.
Unlike in the pure $\kappa$ theory, the growth in the Higgs fields fluctuates even as $\alpha$ increases steadily. The reason for this may be due to the fact that the bands of resonance in Figure \ref{fig:alphamathieu} are much more narrow than in Figure \ref{fig:kappa_mathieu}. As result, the universe may spend very little time in the resonance bands when $\kappa = 0$. Meanwhile, $\phi$ is consistently oscillating as a cosine, and so $\phi$ can accidentally pass through a maximum or minimum as it redshifts through the instability bands. As a result, a small shift in the bands can result in an altered amount of resonance either up or down. But the overall trend for large $\alpha$ is growth. 

Limiting the Higgs growth to be a factor of around 100, as discussed earlier, gives the bound
\begin{equation}
    \alpha < 10^{-8}.
\end{equation}
(This result can be compared Ref.~\cite{Ema:2016kpf} that also placed a bound on this quartic coupling. Related work also includes Refs.~\cite{Espinosa.Riotto.1,Zurek.1,Zurek.2,Herranen:2015ima,Gross:2015bea,Enqvist:2016mqj,Kohri.1,Kohri.2,Jain:2019gsq,Jain:2019wxo,Figueroa:2015rqa,Joti:2017fwe,Ema:2016ehh,Markkanen:2018pdo,Saha:2016ozn,Ema:2017rkk,Gong:2017mwt,Hertzberg:2019prp}.)
Note that here we are only considering the case $\alpha\geq0$, as the negative $\alpha$ case leads to an unbounded inflation-Higgs potential, which we do not consider here. 

\section{Combining Cubic and Quartic}
\begin{figure}[t]
    \centering
    \includegraphics[width=\linewidth]{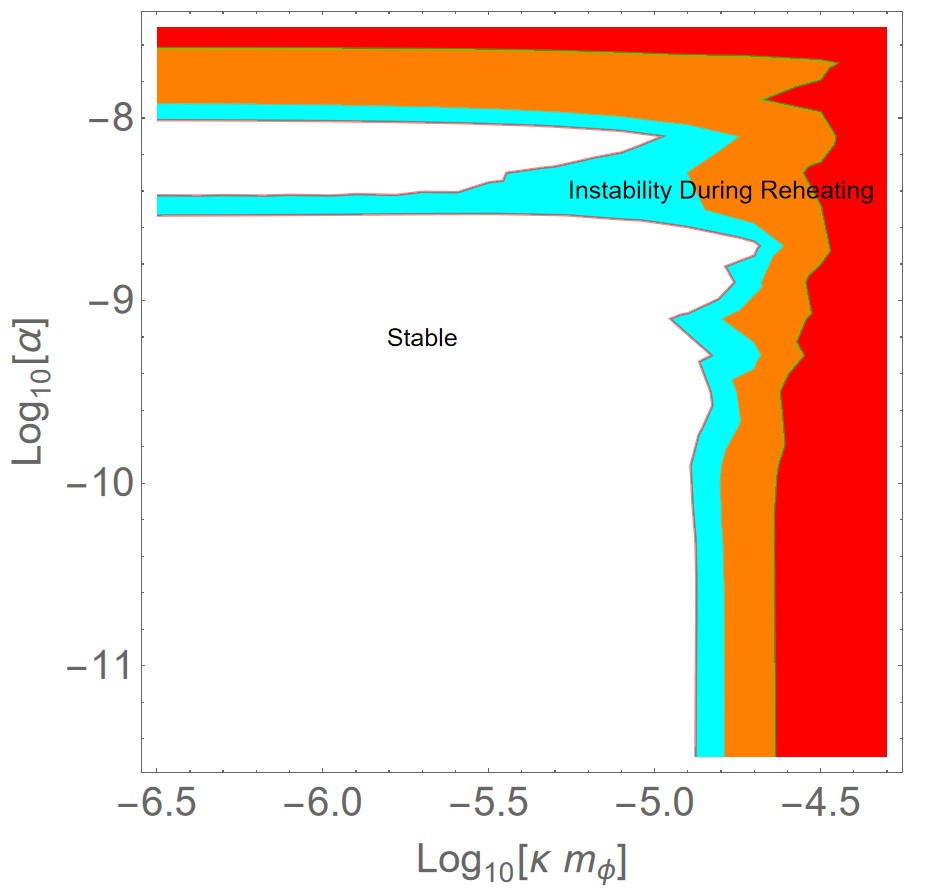}
    \caption{Constraints on $\kappa$ and $\alpha$ during reheating. The colored regions correspond to violating $\langle h^2 \rangle / \langle h^2 \rangle_\tf < 20$ (conservative bound, cyan), 100 (intermediate bound orange), and 3300 (generous bound, red).}
    \label{fig:reheating constraints}
\end{figure}

Finally, the growth in the Higgs field with {\em both} cubic $\kappa$ and quartic $\alpha$ couplings included can be calculated numerically from Eq.~\ref{eq:alphaeom}. The resulting constraints are given in Figure \ref{fig:reheating constraints}. The plot recovers the limits as were previously discussed in this paper, if one sets one of the parameters towards zero. The contour lines again appear jagged, which is due to the same reason the pure $\alpha$ (and to some extent pure $\kappa$) results in Figure \ref{fig:growthalpha} were not smooth. 

Figure \ref{fig:reheating constraints} is one of our primary results, however we shall constrain the allowed parameter space further in the next section.

\section{Instability During Inflation}\label{DuringInflation}
Apart from resonance caused by inflaton couplings during reheating, the Higgs is also at risk of going over its hilltop {\em during} inflation due to the large energy scales at this time. This is because even though the inflaton is slowly rolling during inflation and there is no resonance into the Higgs, there can be important quantum fluctuations. Specifically, the Gibbons-Hawking de Sitter temperature of inflation is \cite{Gibbons:1977mu}
\begin{equation}
    \Tds = \frac{\Hinf}{2\pi}.
\end{equation}
As a result, any light scalar field fluctuates as $\sqrt{\langle h^2 \rangle} \approx \Hinf/2\pi$ per Hubble time. Since inflation lasted about $N_e \approx 55$ Hubble times, the total fluctuation after random walks becomes 
\begin{equation}
    \sqrt{\langle h^2 \rangle} \approx \sqrt{N_e} \frac{H_{end}}{2\pi}
\end{equation}
where $H_{end}$ represents Hubble at the end of inflation and may be close to $\sim 10^{13}$ GeV (it cannot be significantly larger, as this would overproduce gravitational waves). Such fluctuations already exceed the $\hmax$ values seen in Figure \ref{fig:higgsaccurate}. 

\begin{figure}[t]
    \centering
    \includegraphics[width=0.98\linewidth]{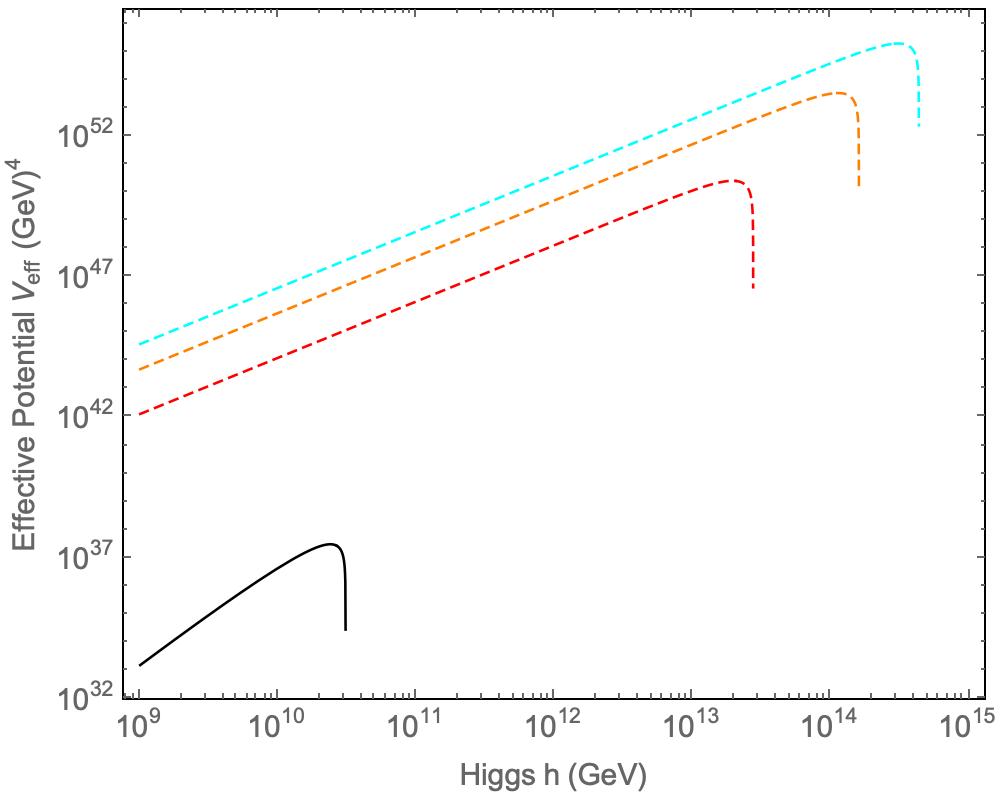}
    \caption{Higgs potential including additional effective masses of 0 (solid black), $1.6\times10^{12}$\,GeV (dashed red), $10^{13}$\,GeV (dashed orange), and $2.8 \times 10^{13}$\,GeV (dashed cyan). This is for the central top mass value of $m_{top}=172.9$\,GeV. The red, orange, and cyan correspond to the generous, intermediate, and conservative limits on the effective mass, respectively.}
    \label{fig:veff}
\end{figure}

Interestingly, coupling to the inflaton can actually help this issue. During inflation, the inflaton field is very large and roughly constant. As a result, the coupling terms in the potential 
$\Delta V = \frac{1}{2}(\kappa\phiinf + \alpha\phiinf^2)h^2$ 
can be interpreted as an additional effective Higgs mass
\begin{equation}
    m_\text{eff}^2 \equiv \kappa\,\phiinf + \alpha\,\phiinf^2.
\end{equation}
 This effective mass can push $\hmax$ to higher energies during inflation and help to avoid causing an instability during inflation. Combining this new effective contribution to the Higgs potential, with the renormalized potential, we obtained the total Higgs potential in Figure \ref{fig:veff}.

In order to save the Higgs from creating an instability, it is necessary that $\hmax^\text{(inf)} \gg \sqrt{\langle h^2 \rangle} \sim 10^{13}$\,GeV. The same probability arguments from earlier can be applied to find the required $\hmax^\text{(inf)}$. For a generous constraint, 
\begin{equation}
    P_\text{inf} \propto \exp\left(-\frac{\left(\hmax^\text{(inf)}\right)^2}{2\langle h^2 \rangle}\right) < e^{-1}
\end{equation}
would put the probability the Higgs growing too large during inflation at around 40\%. This corresponds to 
\begin{equation}
    \hmax^{\text(inf)} > \sqrt{2\langle h^2 \rangle} \approx 2\times 10^{13}\,\mbox{GeV}.
\end{equation}
For a very conservative estimate, the probability of reaching $\hmax^\text{(inf)}$ can be set again to $e^{-3N_e}$ so that only one Hubble patch may become unstable
\begin{equation}
    P_\text{inf} 
    < e^{-3N_e}
\end{equation}
This puts the hilltop at
\begin{equation}
    \hmax^\text{(inf)} > \sqrt{3N_e}\sqrt{2\langle h^2 \rangle} \approx 3\times 10^{14}\,\mbox{GeV}.
\end{equation} 
By adding the effective mass to the Higgs potential, the turnover $\hmax^\text{(inf)}$ can be determined as a function $\kappa$ and $\alpha$. 

The generous constraint corresponds to $m_\text{eff} > m_\text{crit}=1.6 \times 10^{12}$\,GeV while the conservative constraint corresponds to the bound $m_\text{eff} > m_\text{crit}= 2.8 \times 10^{13}$\,GeV. An intermediate value can be taken to be $m_\text{eff} > m_\text{crit}=10^{13}$\,GeV.
Note that towards the end of inflation $\phi$ {\em decreases} to the value $\phiinf$, decreasing $m_\text{eff}$ and therefore this provides the tightest bound. By estimating the end of inflation as $\phiinf = \sqrt{2}\,\Mpl$, the following constraint can be put on the coupling coefficients
\begin{equation}
    \sqrt{2}\,\kappa\, \Mpl + 2\, \alpha\, \Mpl^2 > m_\text{crit}^2.
\end{equation}

\begin{figure}[t]
    \centering
    \includegraphics[width=\linewidth]{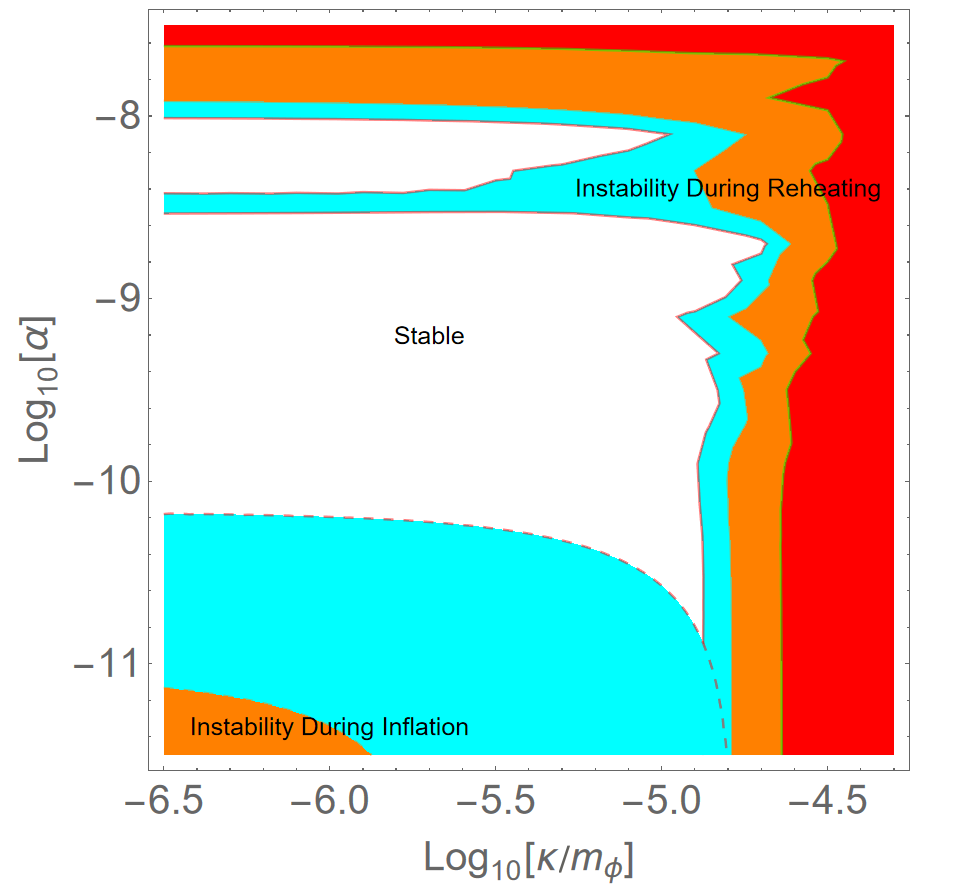}
    \caption{Parameter space due to both reheating and inflation. The (dashed) cyan area corresponds to violating the most conservative constraints $m_\text{eff} > 2.8 \times 10^{13}$\,GeV. The red area corresponding to violating the most generous constraint $m_\text{eff} > 1.6 \times 10^{12}$\,GeV is not contained within the plot shown. The orange area corresponds to violating an intermediate value of $m_\text{eff} > 10^{13}$\,GeV.}
    \label{fig:combined}
\end{figure}

These constraints can be added to Figure \ref{fig:reheating constraints} to produce Figure \ref{fig:combined}, which is our primary result. This figure extends beyond existing results and is useful to summarize the space of allowed couplings.

In this Figure, the white (blank) region is allowed, as the Higgs is sufficiently stable. While the colored regions are ruled out, due to either an instability during reheating (upper right) or an instability during inflation (lower left). Here we focus on $\kappa>0$; for $\kappa<0$, the constraints from inflation are more severe as this coupling worsens the problem (we are adopting the convention that $\phiinf>0$).

Figure \ref{fig:combined} implies that $\alpha$ must be non-zero in order to avoid instabilities during inflation. At the same time, it must be true that $\kappa$ is non-zero in order for inflaton decay to properly reheat the universe. While $\kappa = 0$ would not cause an instability during reheating or during inflation, it would mean that the inflaton could only produce Higgs particles via annihilation. This process requires two inflaton particles to find each other which becomes less and less probable with time.

\section{Conclusion}
In this work, we have taken the SM seriously to high energies. Under this assumption, we have shown that (for weak couplings) the inflaton cubic coupling to the Higgs must have a coupling strength of at most $\kappa < 1.6\times 10^{-5}m_\phi\sim 2.2\times 10^8$\,GeV, while the coefficient for the quartic coupling is constrained by $\alpha < 10^{-8}$. This arises from demanding that the probability of the Higgs field going over its hilltop remains low despite resonance effects. The upper bound on $\kappa$ also places a bound on the the reheat temperature of $\Tr \lesssim 9.2 \times 10^9$ GeV, since in this framework inflaton decays into Higgs would dominate. This may have important implications for classes of models of baryogenesis, which often appeal to extremely high temperatures. Further constraints arise during inflation from de Sitter fluctuations, implying $\alpha$ must be non-zero, while $\kappa$ must also be non-zero for the inflaton to decay efficiently.

Further work includes considering other possible couplings between the inflaton and SM particles. Although these would be higher dimension operators, it would be important to determine under what conditions this may impact the results found here. 

Other directions include the work in Ref.~\cite{Hertzberg:2019prp}, where it was found that  if the inflaton-Higgs cubic coupling $\kappa$ is significantly larger than that focussed on here (namely $\kappa\sim 0.5\,m_\phi$), and if the inflaton is moderately light ($m_\phi\lesssim 10^{12}$\,GeV), one can cure the Higgs potential entirely (this would be in a region far to the right of that displayed in Fig.~\ref{fig:combined}). It would be useful to explore a larger portion of parameter space to consider all these possibilities.

\section*{Acknowledgments}
M.~P.~H is supported in part by National Science Foundation Grant No.~PHY-2013953 and No.~PHY-2310572.

\end{document}